\documentclass{iopconfser}

\usepackage{graphicx}
\usepackage{amsmath}
\usepackage[T1]{fontenc}
\usepackage{lmodern}
\usepackage[colorlinks=true, linkcolor=blue, citecolor=blue, urlcolor=blue]{hyperref}

\begin{document}

\title{Can we identify primordial black holes?\\ The role of subsolar gravitational wave events}

\author{Francesco Crescimbeni$^{1,2}$}

\affil{$^1$Dipartimento di Fisica, Sapienza Università di Roma, Piazzale Aldo Moro 5, 00185, Roma, Italy}
\affil{$^2$INFN, Sezione di Roma, Piazzale Aldo Moro 2, 00185, Roma, Italy}

\email{francesco.crescimbeni@uniroma1.it}

\begin{abstract}
The detection of a subsolar object in a compact binary merger is regarded as one of the most compelling signatures of a population of primordial black holes (PBHs). We critically examine whether such systems can be distinguished from stellar binaries, such as those composed of neutron stars (NSs), which could also populate the subsolar mass range. Unlike PBHs, the gravitational-wave signal from stellar binaries is affected by tidal effects, which increase by several orders of magnitude as the mass decreases. We forecast the capability of current and future gravitational-wave (GW) detectors to constrain tidal effects in putative subsolar binaries. We also discuss the broader implications that the detection of a subsolar merger would have for both cosmology and nuclear physics.
\end{abstract}

\section{Introduction}

The presence of a subsolar-mass (SSM) component in a binary black hole (BBH) system is often cited as the strongest indicator of a primordial origin~\cite{Green:2020jor}. Past LIGO–Virgo–KAGRA (LVK) runs reported SSM candidates, but their low significance prevented their inclusion in the merger catalog~\cite{LIGOScientific:2022hai, KAGRA:2021duu}. Nonetheless, Ref.~\cite{Prunier:2023cyv} analyzed a candidate event—later labeled SSM200308—whose component masses were inferred to be $m_1= 0.62^{+0.46}_{-0.20}M_\odot$ and $m_2= 0.27^{+0.12}_{-0.10} M_\odot$.

However, subsolar masses alone do not uniquely identify PBHs because astrophysical compact objects (e.g., very light neutron stars) could also lie in this range. A crucial discriminant is tidal deformability: material objects are expected to have large tidal responses~\cite{Crescimbeni:2024cwh}, whereas black holes do not, owing to their internal symmetries~\cite{Binnington:2009bb, Damour:2009vw, Chia:2020yla}.

Our goal is to determine, via Bayesian inference, whether tidal measurements can confirm or exclude a PBH origin for an SSM GW signal. We consider both current and next-generation detector networks and discuss the implications of a subsolar merger for fundamental physics.

\section{Subsolar merger modelling}

The lighter the components of a GW binary are, the longer the system takes to merge. Consequently, an SSM binary signal is dominated by the inspiral phase. To model such a signal, we adopt the standard TaylorF2 waveform~\cite{Damour:2000zb}:
\begin{equation}
\tilde{h}(f)=A\, f^{-7/6}\, \exp\left[i\psi(f)\right],
\label{hf}
\end{equation}
where the signal amplitude is proportional to $A\propto{M}_{c}^{5/6}/d_L$, the chirp mass is ${M}_{c}= (m_1 m_2)^{3/5}/(m_1+m_2)^{1/5}$ with $(m_1,m_2)$ the source-frame masses, and $d_L$ the luminosity distance between the GW detector and the binary. The phase is expanded as a power series in the parameter $x=(\pi f M)^{2/3}$, with $f$ the frequency and $M$ the total mass of the binary:
\begin{equation}
\psi(x)=\psi_{\rm pp}(x)+\delta{\psi_{\rm tidal}}(x)\,,
\end{equation}
where $\psi_{\rm pp}(x)$ is the point-particle phase, including spin effects up to 4PN order~\cite{Kidder:1992fr}, and $\delta{\psi_{\rm tidal}}(x)$ is the tidal contribution, containing the 5PN and 6PN tidal terms~\cite{Wade:2014vqa}. Notably, $\delta{\psi_{\rm tidal}}(x)=0$ for PBHs, but is nonzero for any other subsolar compact-object candidate.

\section{Results}

\begin{figure}[ht]
\centering
\includegraphics[width=0.52\textwidth]{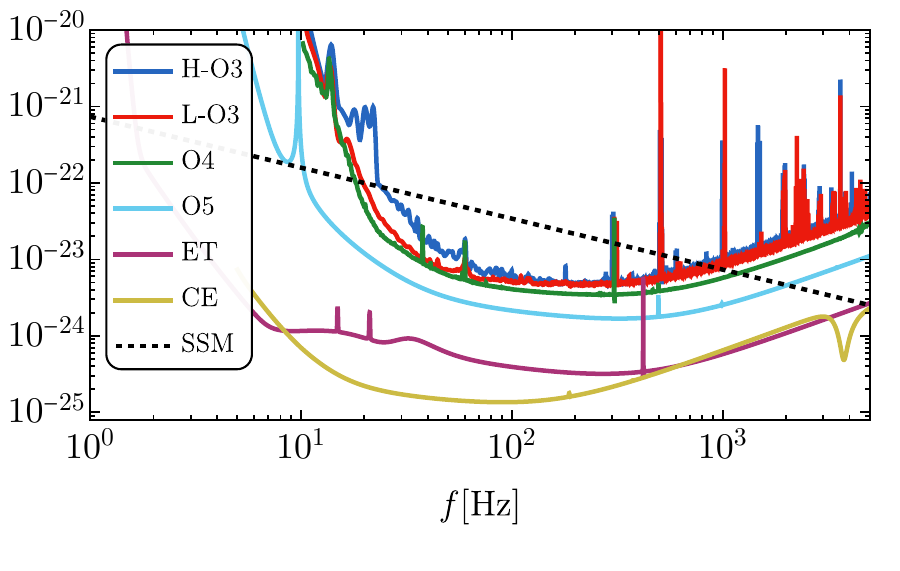}
\caption{ 
Current and future sensitivity curves for the Livingston and Hanford LIGO detectors during the O3, O4, and O5 observing runs, alongside those of the next-generation Einstein Telescope (ET) and Cosmic Explorer (CE). 
The black dashed line indicates the GW amplitude ($2 |\tilde h (f)| \sqrt{f}$), where $\tilde h (f)$ is defined in Eq.~\ref{hf}, for the inspiral phase of an SSM200308-like merger. \textit{Source ref.:} \cite{Crescimbeni:2024cwh}
}\label{fig:sens}
\end{figure}

We perform a mock Bayesian inference analysis using the public software \texttt{BILBY} for an SSM200308-like event, but observed with the ongoing fourth LVK observing run "O4", and with Einstein Telescope plus two Cosmic Explorer detectors (ET+2CE). The injected parameters correspond to those reported for the SSM200308 candidate, with tidal deformabilities set to zero—that is, assuming a PBH origin. Figure~\ref{fig:sens} compares the injected signal with the sensitivity curves of various detectors, showing that third-generation detectors in particular can achieve high sensitivity to such systems.

\begin{figure*}[ht]
\centering
\includegraphics[scale = 0.40]{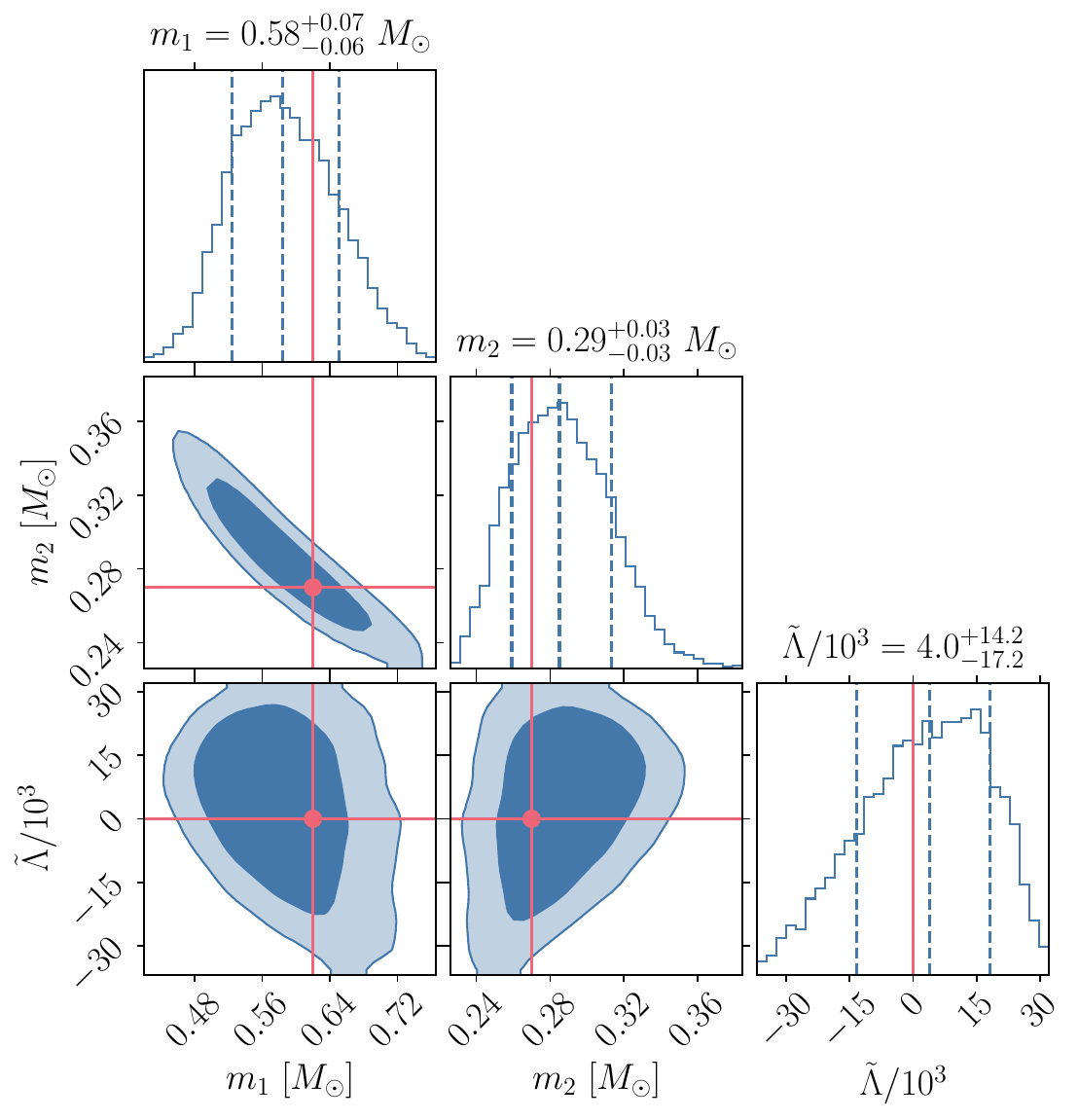} 
\includegraphics[scale = 0.40]{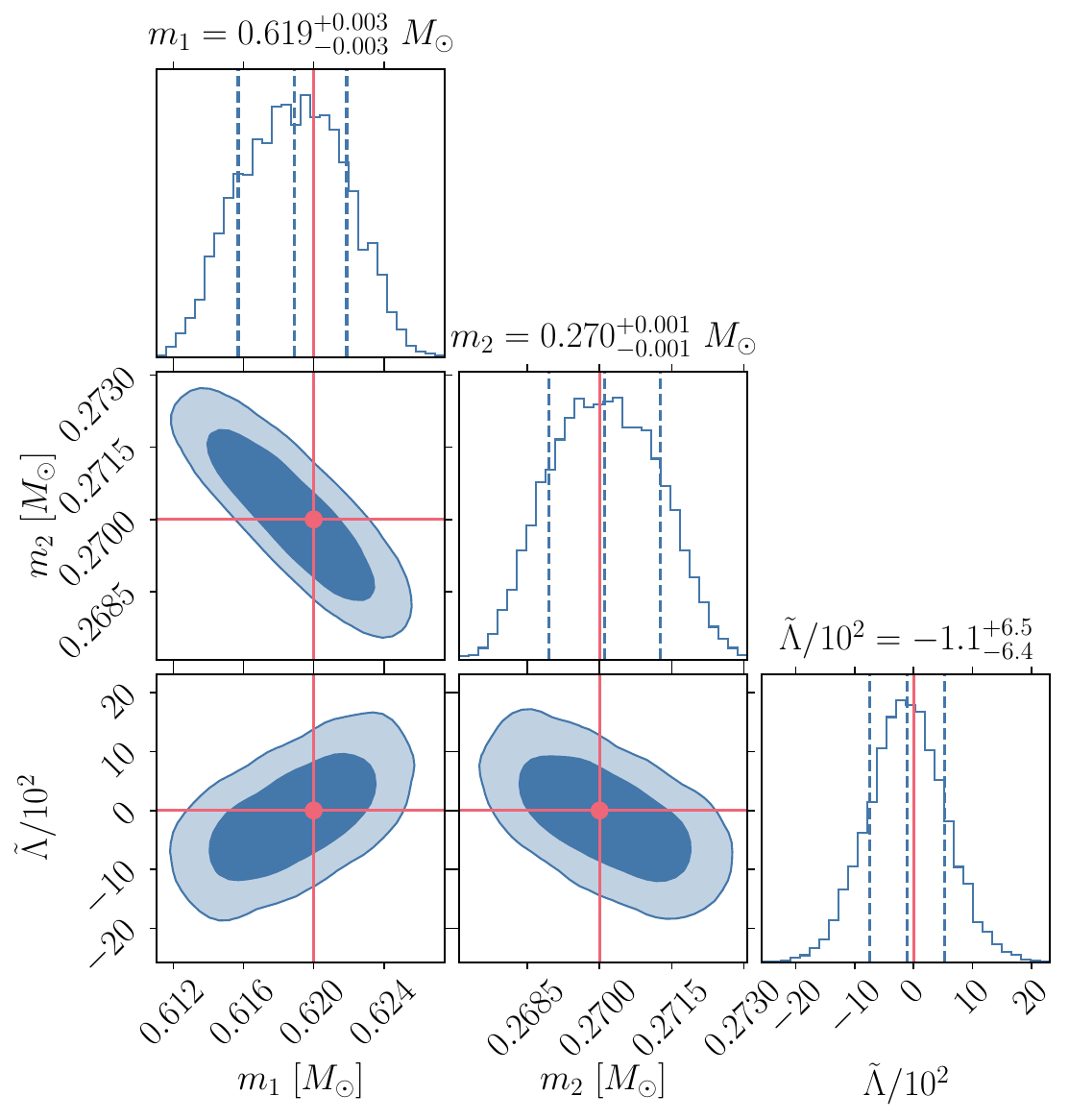}
\caption{Posterior distributions of the source-frame masses $(m_1,m_2)$ and the $\tilde{\Lambda}$ parameter for O4 (left) and ET+2CE (right). The 2D contours correspond to 68$\%$ and 95$\%$ credible intervals. The dashed lines in the 1D histograms indicate 1$\sigma$ intervals relative to the median. Red lines mark the injected values. \textit{Source ref.:} \cite{Crescimbeni:2024cwh}}
\label{fig:corner}
\end{figure*}

We sample on the following parameters:
\begin{equation}\label{eq:paramBBH}
{\boldsymbol \theta}
= \{m_1, m_2, d_L, \theta, \phi, \iota,  \psi, t_c, \Phi_c, \chi_{1}, \chi_{2}, \Lambda_{1}, \Lambda_{2} \}\,, 
\end{equation}
where $\chi_{1,2}$ are the aligned spin magnitudes, $\Lambda_{1}, \Lambda_{2}$ the adimensional tidal deformabilities, $(\theta,\phi)$ the sky position, $\iota$ the binary inclination angle, $\psi$ the polarization angle, $t_c$ the coalescence time, and $\Phi_c$ the coalescence phase. In this way, we can place an upper bound on $\tilde\Lambda$, a combination of $(\Lambda_{1}, \Lambda_{2}, \eta)$, and exclude alternative models predicting large tidal deformabilities. Posterior distributions for $(m_1,m_2,\tilde\Lambda)$ for O4 and ET+2CE are shown in Fig.~\ref{fig:corner}.

These results indicate that to be compatible with the binary neutron star (BNS) hypothesis, an SSM200308-like event would require $\tilde\Lambda\approx10^5$, which is many standard deviations away from the uncertainties shown even in the O4 panel. Therefore, if the injected signal were a PBH binary, the BNS hypothesis could be confidently excluded based on the $\tilde\Lambda$ distribution. In this way, we can confidently identify PBH signatures in a GW signal. Finally, we also find that O4 already provides precise mass measurements, due to the large number of cycles observed in the band.

\section{Implications of a subsolar merger observation}

Having shown that the nature of an SSM can in principle be identified, we now turn to the implications of such a detection.

\subsection{Primordial hypothesis}
A confident SSM BBH-like detection would allow an inference of the PBH abundance $f_{\rm PBH}$—the fraction of the dark matter energy density in PBHs—which controls the expected merger rate. Following Ref.~\cite{Franciolini:2022tfm}, for a model saturating the GWTC-3 upper limit, an abundance of $f_{\rm PBH}\sim{\cal O}(10^{-2})$ could account for an SSM200308-like event, depending on the assumed mass function. This value is close to—but not in conflict with—the strongest OGLE microlensing bounds~\cite{Mroz:2024mse}. Because realistic PBH mass functions should have finite width (e.g., from critical collapse), PBHs might also generate events in the stellar-mass range.

\subsection{NS equation of state}
If SSM objects were instead identified as unusually light neutron stars, this would point to nonstandard formation channels—such as supernovae in dense neutron-rich media~\cite{Metzger:2024ujc}—since conventional supernova models have difficulty producing subsolar NSs~\cite{Muller:2024aod}. The pronounced tidal signatures at these masses would offer powerful constraints on the NS equation of state (EoS). For example, in Ref.~\cite{Crescimbeni:2024qrq} we showed that an SSM detection in O5 could confirm or exclude the existence of quark stars, providing a decisive test of high-density nuclear matter~\cite{Prakash:1995uw, Lattimer:2000nx, Agathos:2015uaa}.

\section{Conclusions}
We have assessed, via Bayesian analysis, whether PBH binaries can be identified and distinguished from other astrophysical compact objects in the subsolar mass range, and at what precision SSM parameters can be measured. First, we showed that O4 already provides precise mass measurements due to the large number of inspiral cycles in the band. Then, we discussed the role of tidal deformabilities. In particular, assuming a PBH origin, we showed that we can constrain tidal deformabilities with an uncertainty of $\approx10^2$ in a 3G detector observation, thus excluding astrophysical candidates like light NSs. Thus, 3G detectors would confidently identify possible signatures of PBHs.

The detection of a subsolar merger would have major implications. If the source is a PBH binary, the measurement would impact cosmology, in particular through the inferred PBH abundance. If the source is a material object, precise measurements of masses, tidal deformabilities, and radii would allow us to place strong constraints on the EoS of dense nuclear matter.

\section*{Acknowledgments}
I would like to thank my collaborators Gabriele Franciolini, Paolo Pani, Antonio Riotto, and Massimo Vaglio. I acknowledge the financial support provided under the ``Progetti per Avvio alla Ricerca Tipo 1,'' protocol number AR12419073C0A82B. Numerical computations were performed at the Vera cluster, supported by MUR and Sapienza University of Rome.

\bibliography{main}

\end{document}